\documentclass[journal]{IEEEtran}
\pdfoutput=1
\usepackage{ifpdf,cite,epsf,psfrag,epsfig}
\usepackage[cmex10]{amsmath}
\usepackage{mathrsfs, amsthm,amsfonts, latexsym, amssymb,algorithmic,algorithm}
\usepackage[mathscr]{eucal}
\usepackage[caption=false,font=normalsize,labelfont=sf,textfont=sf]{subfig}
\usepackage{multicol,multirow,color}
\usepackage[breaklinks]{hyperref}
\usepackage[hyphenbreaks]{breakurl}
\usepackage{makecell}

\usepackage{array}

\begin{document}
\title{Improving the Load Flexibility of Stratified Electric Water Heaters: Design and Experimental Validation of MPC Strategies}

\author{Elizabeth Buechler, Aaron Goldin, and Ram Rajagopal
\thanks{This work was supported in part by DOE ARPA-E under award DE-AR0000697, the Stanford Bits and Watts Initiative, a Stanford Graduate Fellowship, and a NSF Graduate Research Fellowship.}
\thanks{E. Buechler is with the Department of Mechanical Engineering, Stanford University, CA, 94305, USA (email: ebuech@stanford.edu).}
\thanks{A. Goldin is with the Department of Civil and Environmental Engineering, Stanford University, CA, 94305, USA (e-mail: agoldin@stanford.edu).}
\thanks{R. Rajagopal is with the Department of Civil and Environmental Engineering and the Department of Electrical Engineering, Stanford University, CA, 94305, USA (e-mail: ramr@stanford.edu).}
}

\maketitle
\begin{abstract}
Residential electric water heaters have significant load shifting capabilities due to their thermal heat capacity and large energy consumption. Model predictive control (MPC) has been shown to be an effective control strategy to enable water heater load shifting in home energy management systems. In this work, we analyze how modeling tank stratification in an MPC formulation impacts control performance for stratified electric water heaters under time-of-use (TOU) rates. Specifically, we propose an MPC formulation based on a three-node thermal model that captures tank stratification, and compare it to a one-node formulation that does not capture stratification and a standard thermostatic controller. These strategies are compared through both real-time laboratory testing and simulation-based evaluation for different water use patterns. Laboratory experiments show cost reductions of 12.3-23.2\% for the one-node MPC and 31.2-42.5\% for the three-node MPC relative to the thermostatic controller. The performance of the one-node MPC is limited by significant plant-model mismatch, while the three-node formulation better approximates real-world dynamics and results in much more effective cost reduction and load shifting. A simple analysis of how each strategy performs under water use forecast errors is also provided.
\end{abstract}



\section{Introduction}\label{sec:intro}

Demand-side flexibility and energy storage can play an important role in the future integration of variable renewable energy resources into the power grid \cite{ma2016demand}. Behind-the-meter residential loads can provide such flexibility, given appropriate incentives and control strategies. Home energy management systems (HEMS) enable consumers to optimize their electricity consumption in response to time-of-use rates, dynamic prices, demand response requests, or other objectives \cite{zhou2016smart}. Implementing effective HEMS systems requires the development of optimal control strategies for controlling local resources that take into account load dynamics and consumer behavior.

Electric water heaters in particular have significant load shifting capabilities due to their thermal heat capacity, high energy consumption, and high peak power. They currently represent approximately $13.7\%$ of residential electricity consumption in the United States \cite{eia2015residential}, which is likely to increase with residential electrification efforts. Different approaches have been investigated for water heater load management in HEMS systems, including model predictive control (MPC) \cite{jin2017foresee,jin2014model,sossan2013scheduling,shen2021data,cui2019load,blonsky2022home,biagioni2020comparison}, reinforcement learning (RL) \cite{ruelens2016reinforcement,biagioni2020comparison,peirelinck2020domain}, and rule-based control strategies \cite{cui2019load,carew2018heat,vrettos2012load}. Rule-based controllers (e.g., setpoint scheduling) have been demonstrated in laboratory tests and pilot programs for load shifting and peak load reduction \cite{carew2018heat}. However performance may not generalize well to new cost profiles and consumer patterns. Model-free RL generally requires large amounts of training data for good performance \cite{ruelens2016reinforcement}, which may be a barrier to practical adoption. MPC is particularly well suited for water heater control due to its ability to explicitly account for physical constraints, known dynamics, and forecasts of exogenous variables, such as prices and water use patterns. Water heater MPC strategies have been developed for maximizing energy efficiency \cite{jin2014model}, minimizing electricity costs \cite{jin2017foresee,shen2021data}, demand response \cite{jin2017foresee,shen2021data}, and maximizing local PV self-consumption \cite{sossan2013scheduling}. Research has also investigated the impact of forecast uncertainty on performance, and compared deterministic and stochastic MPC strategies \cite{blonsky2022home}.

An important component of an MPC formulation is the control model used to define the tank thermal dynamics. Traditionally, a one-node model which neglects thermal stratification is used \cite{sossan2013scheduling,shen2021data,cui2019load,blonsky2022home,biagioni2020comparison,rosa2023integrating}, resulting in a simple tractable optimization problem when incorporated into an MPC problem. However, most residential water heater tanks are thermally stratified by design to improve energy efficiency \cite{han2009thermal}. A significant amount of literature has focused on developing and validating multi-node ODEs and PDEs for modeling tank stratification \cite{xu2014modeling,han2009thermal}. These models typically have nonlinear dynamics and high state dimension. Therefore, these models are often too complex for use in real-time MPC applications, given that controls must generally be implemented on low-cost microcomputers. The one-node model also cannot capture the effects of multiple control variables and therefore cannot be directly applied to multiple element resistive water heaters or heat pump water heaters with backup resistive elements.

\begin{figure*}\centering
\includegraphics[width=0.90\textwidth]{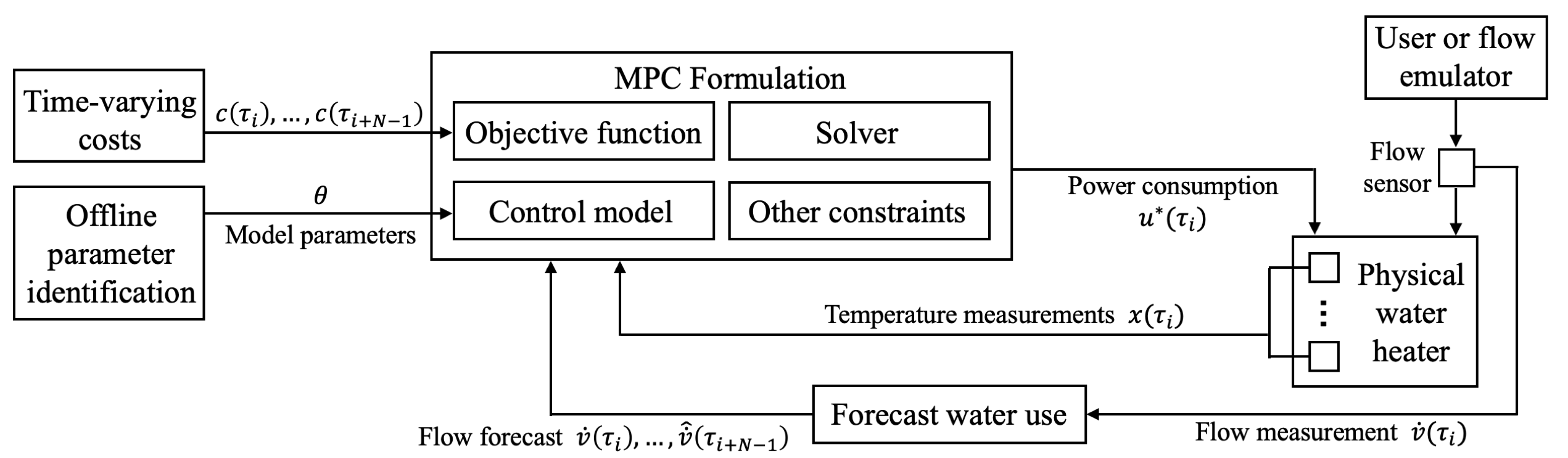}
\caption{Overall control architecture for the water heater MPC strategies.}\label{fig:flow_chart_large}
\end{figure*}

There have been few proposed MPC approaches that use control models that account for tank stratification. Jin et al. \cite{jin2014model,jin2017foresee} and Ritchie et al. \cite{ritchie2021practically} developed optimal control approaches using different two-node models. However, they did not compare their approaches to standard one-node controllers to see whether the use of a stratified model improves control performance. Awadelrahman et al. \cite{awadelrahman2017economic} and Kepplinger et al. \cite{kepplinger2014demand} developed optimal control formulations using higher-order multi-node models. These proposed optimization problems are nonconvex and may be too computationally intensive to be reasonably implemented on a low-cost microcomputer. The paper by Kepplinger et al. \cite{kepplinger2014demand} is the only known prior work that has investigated whether the use of a higher fidelity control model improves control performance. Results showed that the control formulation using the nonlinear multi-node thermal model performed better than the controller using a one-node model.

A limitation of these prior works is that controller performance has been evaluated mainly in simulation. The accuracy of such studies depends heavily on how the ground-truth system is modeled. Even PDE and higher-order multi-node ODE models do not perfectly capture real-world tank dynamics and must be carefully tuned to measured data. While some research has conducted lab testing of optimal control strategies\cite{baumann2023experimental,halamay2019hardware,kepplinger2016field,starke2020real}, the focus has mainly been on approaches using one-node control models. Additional laboratory or field-based testing is needed to validate controller performance under realistic conditions, and understand how control model fidelity affects MPC performance.

In this work, we analyze how modeling tank stratification in an MPC formulation impacts control performance for stratified electric water heaters under time-varying electricity prices, using both experimental testing and simulation-based analysis. We also propose a new formulation which achieves improved performance while remaining computationally tractable. Our contributions can be summarized as follows:
\begin{itemize}
    \item We propose a novel water heater MPC formulation designed to capture the stratification patterns observed in two-element resistive water heaters. The optimization formulation is convex and computationally tractable and uses a linear three-node model of tank thermal dynamics. The sensing, communications, hardware, and computational requirements of the controller were carefully designed to ensure that it is physically implementable on real-world water heaters.
    \item We develop a parameter identification method for the proposed three-node controller model and validate it using real-world measurements.
    \item We design a laboratory testbed with the sensing and measurement capabilities to validate different control strategies under realistic conditions.
    \item Extensive real-time tests of the controller were performed using the laboratory testbed and in simulation to compare the proposed controller to a one-node MPC strategy and a thermostatic controller. Performance was evaluated under different water use patterns and water forecast errors.
\end{itemize}

The paper is organized as follows: In Section \ref{sec:mpc} we describe the MPC formulations and methods for controller model parameter identification. In Section \ref{sec:valid} we describe the laboratory and simulation-based testbeds used for validating control performance. Case study results on parameter identification and controller performance are presented in Section \ref{sec:case_studies} and conclusions and future work are discussed in Section \ref{sec:con}.

\section{Water Heater Modeling and Control}\label{sec:mpc}

In this work, we focus on two-element resistive water heaters, which are very common in the United States and other countries. We also focus on water heaters without thermostatic mixing valves (e.g., hot water is not mixed to a lower temperature before distribution).

MPC is an optimal control strategy which involves solving an open-loop optimization problem to find control actions which minimize a cost function, subject to constraints on the system and predictions of future disturbances and exogenous variables \cite{borrelli2017predictive}. At each iteration, only the first computed control action is actuated on the physical system before re-measuring the state and re-solving the optimization at the next iteration to close the loop.

For water heater control, the objective is to minimize electricity costs under time-varying prices while keeping the water temperature within an acceptable range to maintain comfort requirements. We assume consumers are subject to a time-of-use (TOU) retail rate $c(\tau_j)$ $[\$/kWh]$. A control model is used to represent the tank thermal dynamics in the optimization problem. Parameters of the control model are estimated offline using measured data. For the MPC strategies considered, the state variables $x(\tau_i)$ are measured tank temperature(s), the control variables $u(\tau_i)$ are the power consumption of the element(s), and the disturbance is the water flow rate. The specific variable definitions for each specific MPC formulation are described in Sections \ref{sec:1node} and \ref{sec:3node} and are listed in Table \ref{tab:variable_def}.

\begin{table}[]
\caption{State and control variables for MPC formulations}
    \centering
    \begin{tabular}{|c|c|c|c|}
      \hline
      Variable   & \makecell{Physical\\representation} & \makecell{One-node\\MPC} & \makecell{Three-node\\MPC} \\
      \hline
       State $x$ &Temperature(s) [K] & $T$ & [$T_l$, $T_m$, $T_u$] \\
       Control $u$ & Element power(s) [W] & $p$ & $[p_m,p_u]$\\
       Disturbance & Flow rate [$m^3/s$] & $\dot{v}$ & $\dot{v}$\\
       \hline
    \end{tabular}
    \label{tab:variable_def}
\end{table}

The MPC is implemented in a receding horizon manner over a control horizon $T_N$ from current time $\tau_i$ to $\tau_i+T_N$. The control horizon is discretized into $N$ control intervals of length $\delta t$, such that $N=T_N/\delta t$. Timesteps are indexed such that $\tau_j=\tau_i+(j-i) \delta t$. The MPC algorithm proceeds as follows: at time $\tau_i$, the temperature state $x(\tau_i)$ and the water flow rate $\dot{v}(\tau_i)$ are measured. Then the future flow rate is forecast over the control horizon, producing estimates $[\hat{\dot{v}}(\tau_i),\dots,\hat{\dot{v}}(\tau_{i+N-1})]$. The MPC problem is then solved, producing optimal temperature state trajectory $\mathbf{x}^*=[x^*(\tau_i),\dots,x^*(\tau_{i+N})]$ and control action (power consumption) trajectory $\mathbf{u}^*=[u^*(\tau_i),\dots,u^*(\tau_{i+N-1})]$. The first control action $u^*(\tau_i)$ is actuated on the physical system. Then the algorithm is repeated at time $\tau_{i+1}$. The overall architecture is shown in Fig. \ref{fig:flow_chart_large}.

In the following subsections we describe MPC optimization problems based on one-node and three-node control models and parameter identification methods for the control models.

\subsection{One-node MPC}\label{sec:1node}

\begin{figure}
\centering
\includegraphics[width=0.49\textwidth]{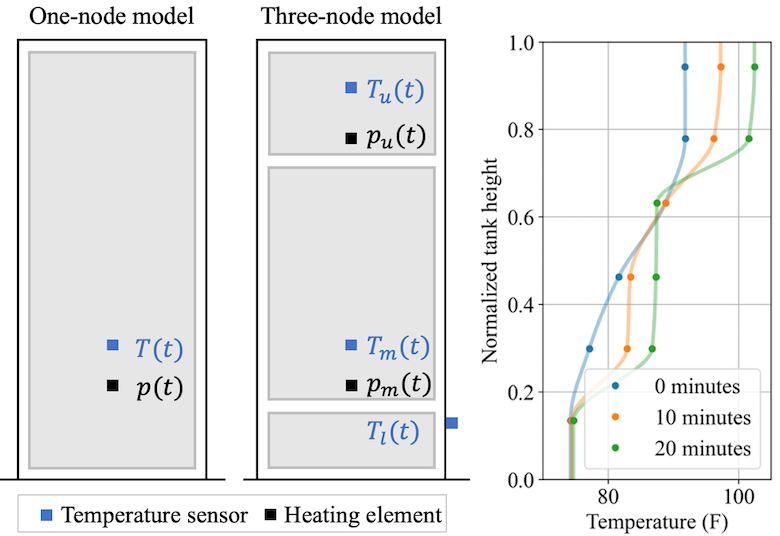}
\caption{Depiction of the one-node and three-node water heater thermal models, with heating element locations and temperature sensor placements. The plot on the right shows tank temperature profile measurements from an actual water heater after both heating elements have been on for 0, 10, and 20 minutes, showing the development of three well-mixed regions. The line plots are a spline interpolation of the measurements and are used for visualization.}\label{fig:model_diagram}
\end{figure}

The one-node thermal model assumes that the tank is fully mixed at a uniform temperature $T(\tau_j)$. Since it cannot capture the affects of multiple control variables on tank thermal dynamics, we assume that if a one-node MPC strategy were applied to a two-element water heater, that only the lower element would be used. The state and control variables for this formulation are therefore defined as $x(\tau_j)=T(\tau_j)$ and $u(\tau_j)=p(\tau_j)$, respectively, where $p(\tau_j)$ is the average power [W] of the lower heating element from $\tau_j$ to $\tau_{j+1}$. The temperature $T(\tau_j)$ $[K]$ is measured using a single sensor located right above the lower element, where a thermostat is normally located (Fig. \ref{fig:model_diagram}).

At time $\tau_i$, the following optimization problem is solved:
\begingroup
\allowdisplaybreaks
\begin{subequations}
\begin{align}
    &\underset{\mathbf{T},\mathbf{p}}{\text{minimize}}\quad \sum_{j=i}^{i+N-1}{\left[J_{C_j}(p(\tau_j))+\lambda J_{T_j}(T(\tau_j))\right]}\\
    &\text{subject } \text{to:}\nonumber\\  &\qquad T(\tau_{j+1})=f(T(\tau_{j}),p(\tau_{j}))\qquad\quad\:\: \forall j\in\mathcal{J}\label{eq:1node_dyn}\\
    &\qquad T(\tau_i)=\tilde{T}(\tau_i)\label{eq:1node_init}\\
    &\qquad 0\leq p(\tau_j)\leq \bar{p} \:\: \:\qquad\qquad\qquad\qquad \forall j\in\mathcal{J}\label{eq:1node_power}
\end{align}
\end{subequations}
\endgroup

where  $\mathcal{J}=\{i,\dots,i+N-1\}$, $\mathbf{T}=[T(\tau_i),\dots,T(\tau_{i+N})]$, and $\mathbf{p}=[p(\tau_i),\dots,p(\tau_{i+N-1})]$. $\lambda$ is a weighting factor. The objective function includes an electricity cost term $J_{C_j}(p(\tau_j))$ and a temperature penalty term $J_{T_j}(T(\tau_j))$. The electricity cost $[\$]$ is given by:
\begin{align}
    J_{C_j}(p(\tau_j))=\frac{\delta t}{3.6e6}c(\tau_j)p(\tau_j)
\end{align}
The factor in the denominator accounts for unit conversion. The temperature term is a convex deadband quadratic function that penalizes deviations outside of a specified temperature range:
\begin{align}
    J_{T_j}(T(\tau_j))=\left[\underline{T}-T(\tau_j)\right]_{+}^2+\beta\left[T(\tau_j)-\bar{T}\right]_{+}^2
\end{align}
where the notation $[\cdot]_+$ is equivalent to $\text{max}(0,\cdot)$. $\underline{T}$ and $\bar{T}$ are soft lower and upper bounds on the tank temperatures, and $\beta$ is a weighting factor. A soft temperature penalty is used instead of a hard constraint to ensure that there is always a feasible solution. 

Equation (\ref{eq:1node_dyn}) defines the system dynamics in discrete time. In continuous time, the one-node model is given by: 
\begin{align}\label{eq:1node_cont}
        \frac{dT(t)}{dt}=\frac{p(t)}{C}+\frac{U}{C}(T_a-T(t))-\frac{\dot{v}(t)}{V}(T(t)-T_{i})
\end{align}
where $T_a$ is the ambient air temperature $[K]$, $T_{i}$ is the inlet water temperature $[K]$, $\dot{v}$ is the volumetric water flow rate $[m^3/s]$, $C$ is the thermal capacitance of the tank $[J/K]$ and $U$ is the thermal conductance of the tank insulation $[W/K]$. The thermal capacitance is defined as $C=c_p \rho V$, where $c_p$ is the specific heat capacity of water $[J/kg\cdot K]$, $\rho$ is the density of water $[kg/m^3]$, and $V$ is the volume of the tank~$[m^3]$. We assume that $T_a$ is measured by the HEMS system and $T_i$ is known or can be estimated, and both are constant over time. Since we focus on water heaters without mixing valves, we assume that the volumetric flow rate $\dot{v}$ out of the tank is not dependent on the water temperature. This is a valid assumption given that the range of acceptable outlet water temperatures is relatively small. Equation (\ref{eq:1node_cont}) can be converted to a difference equation using an explicit numerical timestepping method (e.g., forward Euler), resulting in Equation (\ref{eq:1node_dyn}), which is an affine equality constraint.

Equation (\ref{eq:1node_init}) defines the initial condition, where $\tilde{T}(\tau_i)$ is the measured temperature at time $\tau_i$. Equation (\ref{eq:1node_power}) constrains the heating element power to be non-negative and less than or equal to the nominal power rating $\bar{p}$. For most resistive water heaters, power is controlled via ON/OFF switching of the element. However, some water heaters can continuously modulate element power by controlling the voltage \cite{yildiz2021assessment}. The proposed MPC can accommodate both approaches. If power modulation is possible, $p(\tau_j)$ represents the actual (constant) power consumption in the control interval. Otherwise, it represents the average power and it is converted to an ON/OFF signal for implementation. Unlike heat pump water heaters, resistive water heaters do not have short-cycling constraints, so converting an average power to an ON/OFF signal is straightforward. This approach allows us to avoid including integer constraints in the formulation.

\subsection{Three-node MPC} \label{sec:3node}

In this section we propose an MPC strategy based on a three-node thermal model that coarsely accounts for tank stratification while remaining computationally tractable. The model assumes there are three volumes of water, each with a different temperature (Fig. \ref{fig:model_diagram}). This model was chosen based on the empirical observation that under many operating conditions, two-element resistive water heaters have a temperature profile with three distinct, relatively well-mixed regions: (i) above the upper element, (ii) between the two elements, and (iii) below the lower element. These regions tend to be well-mixed if the elements have recently been on for a period of time, due to buoyancy effects. Fig. \ref{fig:model_diagram} shows  measurements from an actual water heater, demonstrating how the tank temperature profile evolves after the two elements are turned on, starting from an arbitrary profile. After 10-20 min, the temperature in each region becomes more uniform. In other operating conditions, such as after large water draws or after long periods with no power consumption, the temperatures become less uniform. 

The state variables for this MPC formulation include the temperatures of the lower, middle, and upper nodes $x(\tau_j)=[T_l(\tau_j),T_m(\tau_j),T_u(\tau_j)]^T$, and the control variables include the power consumption of the two elements $u(\tau_j)=[p_m(\tau_j),p_u(\tau_j)]^T$. We assume the upper and middle node temperatures are measured with sensors right above each element, where thermostats are typically located (Fig. \ref{fig:model_diagram}). The temperature of the lower node is measured with a sensor below the lower element. 

The following optimization problem is solved at time $\tau_i$:
\begingroup
\allowdisplaybreaks
\begin{subequations}
\begin{align}
    &\underset{\substack{\mathbf{T_l},\mathbf{T_m},\mathbf{T_u}\\\mathbf{p_m},\mathbf{p_u}}}{\text{minimize}} \sum_{j=i}^{i+N-1}{\left[J_{C_j}(p_m(\tau_j),p_u(\tau_j))+\lambda J_{T_j}(T_u(\tau_j))\right]}\\
    &\text{subject to:}\nonumber\\
    &\:\:\: T_u(\tau_{j+1})=f_u(T_u(\tau_j), T_m(\tau_j),p_u(\tau_j))\qquad\forall j\in\mathcal{J}\label{eq:3node_dyn1}\\
    &\:\:\:T_m(\tau_{j+1})=f_m(T_u(\tau_j), T_m(\tau_j),T_l(\tau_j), p_m(\tau_j))\nonumber\\
    &\qquad\qquad\qquad\qquad\qquad\qquad\qquad\qquad\qquad\:\:\:\:\forall j\in\mathcal{J}\label{eq:3node_dyn2}\\
    &\:\:\:T_l(\tau_{j+1})=f_l(T_m(\tau_j), T_l(\tau_j))\qquad\qquad\quad\:\:\:\: \forall j \in \mathcal{J}\label{eq:3node_dyn3}\\
    &\:\:\:T_l(\tau_j)\leq T_m(\tau_j)\qquad\qquad\qquad\qquad\qquad\quad\:\:\forall j \in \mathcal{K}\label{eq:3node_buo1}\\
    &\:\:\:T_m(\tau_j)\leq T_u(\tau_j)\qquad\qquad\qquad\qquad\qquad\quad\: \forall j \in \mathcal{K}\label{eq:3node_buo2}\\
    &\:\:\:0\leq p_m(\tau_j)\leq \bar{p}_m\qquad\qquad\qquad\qquad\qquad\:\:\:\: \forall j\in\mathcal{J}\label{eq:3node_pow1}\\
    &\:\:\:0\leq p_u(\tau_j)\leq \bar{p}_u\qquad\qquad\qquad\qquad\qquad\quad\:\: \forall j\in\mathcal{J}\label{eq:3node_pow2}\\
    &\:\:\:T_u(\tau_i)=\tilde{T}_u(\tau_i)\label{eq:3node_init1}\\
    &\:\:\:T_m(\tau_i)=\tilde{T}_m(\tau_i)\label{eq:3node_init2}\\
    &\:\:\:T_l(\tau_i)=\tilde{T}_l(\tau_i)\label{eq:3node_init3}
\end{align}
\end{subequations}
\endgroup
where $\mathbf{T_x}=[T_x(\tau_i),\dots,T_x(\tau_{i+N})]$ for $x\in\{l,m,u\}$, $\mathbf{p_x}=[p_x(\tau_i),\dots,p_x(\tau_{i+N-1})]$ for $x\in\{m,u\}$, $\mathcal{J}=\{i,\dots,i+N-1\}$, $\mathcal{K}=\{i,\dots,i+N\}$, and $\lambda$ is a weighting factor. 

The objective function includes a temperature penalty term and an electricity cost term. The temperature term is a function of the temperature of the upper node (near the outlet):
\begin{align}
    J_{T_j}(T_u(\tau_j))=\left[\underline{T}-T_u(\tau_j)\right]_{+}^2+\beta\left[T_u(\tau_j)-\bar{T}\right]_{+}^2
\end{align}
The electricity cost is a function of the power consumption of both the upper and lower elements $p_u(\tau_j)$ and $p_m(\tau_j)$:
\begin{align}
    J_{C_j}(p_m(\tau_j),p_u(\tau_j))=\frac{\delta t}{3.6e6}\,c(\tau_j)\,(p_m(\tau_j)+p_u(\tau_j))
\end{align}
Equations (\ref{eq:3node_dyn1})-(\ref{eq:3node_dyn3}) define the system dynamics in discrete time. In continuous time, the temperature dynamics of the upper node are modeled as:
{\small
\begin{align}
    \frac{dT_u(t)}{dt}&=\frac{U_u}{C_u }\,(T_a-T_u(t))-\frac{\dot{v}(t)}{V_u}\,(T_u(t)-T_m(t))\nonumber\\& +\frac{p_u(t)}{C_u}+\frac{K_{um}}{C_u}\,(T_m(t)-T_u(t))\label{eq:dyn1}
\end{align}}
where $U_u$ is the thermal conductance of the tank insulation of the upper node. $C_u$ is the thermal capacitance of the upper node, which is defined as $C_u=V_u\rho c_p$, where $V_u$ is the node volume. $K_{um}$ is the modeled lumped thermal conductivity between the upper and middle nodes. For the middle node, the temperature dynamics are given by:
{\small
\begin{align}
    \frac{dT_m(t)}{dt}&=\frac{U_m}{C_m }\,(T_a-T_m(t))-\frac{\dot{v}(t)}{V_m}\,(T_m(t)-T_l(t))+\frac{p_m(t)}{C_m} \nonumber\\& +\frac{K_{ml}}{C_m}\,(T_l(t)-T_m(t))+\frac{K_{um}}{C_m}\,(T_u(t)-T_m(t))\label{eq:dyn2}
\end{align}}
Similarly for the lower node, the dynamics are defined as:
{\small
\begin{align}
    \frac{dT_l(t)}{dt}&=\frac{U_l}{C_l}\,(T_a-T_l(t))-\frac{\dot{v}(t)}{V_l}\,(T_l(t)-T_i)\quad\quad\quad\nonumber\\&+\frac{K_{ml}}{C_l}\,(T_m(t)-T_l(t))\label{eq:dyn3}
\end{align}}
This model has similarities to others proposed in the literature, such as \cite{zuniga2017parameter}. Equations (\ref{eq:dyn1})-(\ref{eq:dyn3}) account for ambient losses, uniform flow from the bottom to the top of the tank, conduction between adjacent nodes, and heating by the two elements. The model does not explicitly include nonlinear buoyancy terms, which typically affect dynamics in lumped parameter or multi-node water heater tank models when there are temperature inversions between neighboring nodes. Instead, we include constraints in the optimization problem which restrict the state to the region where temperature inversions do not occur (Equations (\ref{eq:3node_buo1}) and (\ref{eq:3node_buo2})), similar to the approach taken in \cite{jin2017foresee}. We found through simulation and laboratory testing that these additional constraints are in practice not very restrictive, as it is generally optimal to prioritize the use of the upper element.

The continuous dynamics given in Equations (\ref{eq:dyn1})-(\ref{eq:dyn3}) can be written as difference equations by applying an explicit numerical method (e.g., forward Euler), resulting in Equations (\ref{eq:3node_dyn1})-(\ref{eq:3node_dyn3}), which are affine equality constraints. Functions $f_u(\cdot)$, $f_m(\cdot)$, and $f_l(\cdot)$ are defined by Equations (\ref{eq:dyn1})-(\ref{eq:dyn3}) and the specific numerical method.

Equations (\ref{eq:3node_pow1}) and (\ref{eq:3node_pow2}) are constraints on the average power of the heating elements. Equations (\ref{eq:3node_init1})-(\ref{eq:3node_init3}) ensure that the initial conditions hold, where $\tilde{T}_u(\tau_i)$, $\tilde{T}_m(\tau_i)$, and $\tilde{T}_l(\tau_i)$ are temperature measurements at time $\tau_i$. In practice, it is possible for the initial conditions to slightly violate Equations (\ref{eq:3node_buo1}) and (\ref{eq:3node_buo2}) due to sensor noise, calibration error, or other unmodeled effects. In such cases the initial conditions must be adjusted before solving the optimization problem to ensure feasibility.

\subsection{Control Implementation}

Both the one-node and three-node MPC formulations are convex optimization problems. We use the CasADi python optimization package \cite{andersson2019casadi} with the IPOPT solver \cite{wachter2006implementation} to solve each problem.

The performance of the MPC strategies was compared to a baseline thermostatic controller that is often used in two-element resistive water heaters. The elements are each operated in a hysteresis deadband between $\underline{T}$ and $\bar{T}$. They are operated non-simultaneously, with the upper element given priority if the temperature of the upper element falls below $\underline{T}$. The temperature sensors associated with each element are located right above the element.

\subsection{Parameter Identification}\label{sec:param_id}

The one-node and three-node thermal models are both physics-based grey-box models, where some physical constants are known, and other design-specific parameters can be estimated from measurements. We propose methods for parameter identification of both models using ordinary least squares (OLS) regression. In Section \ref{sec:param_id_results} we show that physically realistic parameter estimates are obtained.

\subsubsection{One-node model}

For the one-node model, known constants are the density of water ($\rho=1000$ $[kg/m^3]$) and the thermal heat capacity of water ($c_p=4181.3$ $[J/kg\cdot K]$). Unknown parameters are the tank volume $V$ and the thermal conductance of the tank insulation $U$. For parameter estimation, we let $\dot{v}=0$ and discretize the continuous system dynamics (Equation \ref{eq:1node_cont}) using forward Euler timestepping:
\begin{align}
    T(\tau_{j+1})=T(\tau_j)+\bar{\delta}t\left[\frac{U}{\rho c_p V}(T_a-T(\tau_j))+\frac{p(\tau_j)}{\rho c_p V}\right]\label{eq:1node_euler}
\end{align}
$\bar{\delta}t$ is the integration timestep of the forward Euler method, which is equal to $\bar{\delta t}=\delta t/m$, where $m$ is the number of forward Euler steps in the control interval. Equation (\ref{eq:1node_euler}) can be expressed as an energy balance for the tank:
\begin{align}
    \bar{\delta} t p(t)= \rho c_p V (T(\tau_{j+1})-T(\tau_j))+U\bar{\delta}t (T(\tau_j)-T_a)
\end{align}
The left-hand term represents the energy [J] consumed by the element between time $\tau_j$ and $\tau_{j+1}$, the first right-hand term is the change in energy in the tank between timesteps, and the last term is the ambient loss. This expression is a linear function of the parameters $\theta=[V,\: U]^T$, and can be written as $z_j=w_j^T \theta $, where $w_j\in\mathbb{R}^2$ and $z_j\in\mathbb{R}$ are both a function of known or measured variables:
\begin{align}
    z_j&=\bar{\delta} t p(\tau_j)\\
    w_j&=[\rho c_p (T(\tau_{j+1})-T(\tau_j)),\:\bar{\delta}t(T(\tau_j)-T_a)]^T
\end{align}
Given a dataset of measured variables $T(\tau_j)$, $T(\tau_{j+1})$, $p(\tau_j)$, and $T_a$, we solve for $\theta$ via linear regression such that the parameter values minimize the sum of squared residuals of the energy balance in the tank.

The linear regression is performed using data from two different operating conditions: (i) during a heating cycle (lower element is on and no water flow), and (ii) when the water heater is at rest (no water flow and no power). We find that using data collected under these two simple operating conditions for parameter identification generally results in reliable, physically realistic parameter estimates, as long as the assumptions of the one-node model hold when the data is collected. This approach is validated in Section \ref{sec:param_id_results}.

\subsubsection{Three-node model}
For the three-node model, it is assumed that $\rho$ and $c_p$ are known, and the total tank volume $V$ is known from manufacturer tank specifications. The lower volume $V_l$ is expressed as a function of the other variables such that $V_l=V-V_m-V_u$. Similar to the one-node case, we let $\dot{v}=0$ and discretize the continuous system dynamics defined in Equations (\ref{eq:dyn1})-(\ref{eq:dyn3}) using forward Euler timestepping and rewrite each expression as an energy balance for each node:
\begin{align}
&\bar{\delta} t p_u(\tau_j)=U_u \bar{\delta} t (T_u(\tau_j)-T_a)+K_{um}\bar{\delta}t(T_u(\tau_j)-T_m(\tau_j))\nonumber\\
&\quad\quad\quad\quad +V_u\rho c_p(T_u(\tau_{j+1})-T_u(\tau_j))\\
&\bar{\delta}t p_m(\tau_j)=U_m\bar{\delta}t(T_m(\tau_j)-T_a)+K_{ml}\bar{\delta}t(T_m(\tau_j)-T_l(\tau_j))\nonumber\\
&\quad\quad\quad\quad +K_{um}\bar{\delta}t(T_m(\tau_j)-T_u(\tau_j))\nonumber\\
&\quad\quad\quad\quad +V_m\rho c_p (T_m(\tau_{j+1})-T_m(\tau_j))\\
&V\rho c_p(T_l(\tau_{j})-T_l(\tau_{j+1}))=U_l\bar{\delta}t(T_l(\tau_j)-T_a)\nonumber\\
&\quad\quad\quad\quad +K_{ml}\bar{\delta}t(T_l(\tau_j)-T_m(\tau_j))\nonumber\\
&\quad\quad\quad\quad -(V_m+V_u)\rho c_p(T_l(\tau_{j+1})-T_l(\tau_j))
\end{align}
These expressions are linear functions of the parameters $\theta =[U_l,\:U_m,\:U_u,\:K_{ml},\:K_{um},\:V_m,\:V_u]^T\in\mathbb{R}^7$ and can written as $z_j=W_j\theta$, where $z_j\in\mathbb{R}^3$ and $W_j\in\mathbb{R}^{3\times7}$ are a function of known or measured variables:
\begin{align}
    z_j=[\bar{\delta}t p_u(\tau_j),\bar{\delta}t p_m(\tau_j),V \rho c_p (T_l(\tau_{j})-T_l(\tau_{j+1}))]^T
\end{align}
The nonzero elements of $W_j$ are defined as:
\begingroup
\allowdisplaybreaks
\begin{align}
    W_{j,13}=&\bar{\delta }t(T_u(\tau_j)-T_a)\\
    W_{j,15}=&\bar{\delta }t(T_u(\tau_j)-T_m(\tau_j))\\
    W_{j,17}=&\rho c_p (T_u(\tau_{j+1})-T_u(\tau_j))\\
    W_{j,22}=&\bar{\delta}t(T_m(\tau_j)-T_a)\\
    W_{j,24}=&\bar{\delta} t(T_m(\tau_j)-T_l(\tau_j))\\
    W_{j,25}=&\bar{\delta}t(T_m(\tau_j)-T_u(\tau_j))\\
    W_{j,26}=&\rho c_p (T_m(\tau_{j+1})-T_m(\tau_{j}))\\
    W_{j,31}=&\bar{\delta}t(T_l(\tau_j)-T_a)\\
    W_{j,34}=&\bar{\delta}t(T_l(\tau_j)-T_m(\tau_j))\\
    W_{j,36}=&-\rho c_p (T_l(\tau_{j+1})-T_l(\tau_{j}))\\
    W_{j,37}=&-\rho c_p (T_l(\tau_{j+1})-T_l(\tau_{j}))
\end{align}
\endgroup
where $W_{j,ik}$ is the element in the $i$th row and $k$th column of the matrix $W_j$. Given a dataset of measured variables, we solve for $\theta$ via linear regression such that the parameter values minimize the sum of squared residuals of the nodal energy balances. Datasets were collected under the same two operating conditions (heating cycle and at rest) as for the one-node model. However for the heating cycle, both elements were turned on at the same time. This method is validated in Section \ref{sec:param_id_results}.

\section{Laboratory and Simulation Testbeds}\label{sec:valid}

The performance of the controllers was evaluated in laboratory tests using an actual water heater and in simulation using a high-fidelity water heater tank model. These laboratory and simulation-based testbeds are described in the following subsections.

\subsection{Laboratory Water Heater Testbed}\label{sec:valid_exp}

\begin{figure}
\centering
\includegraphics[width=0.48\textwidth]{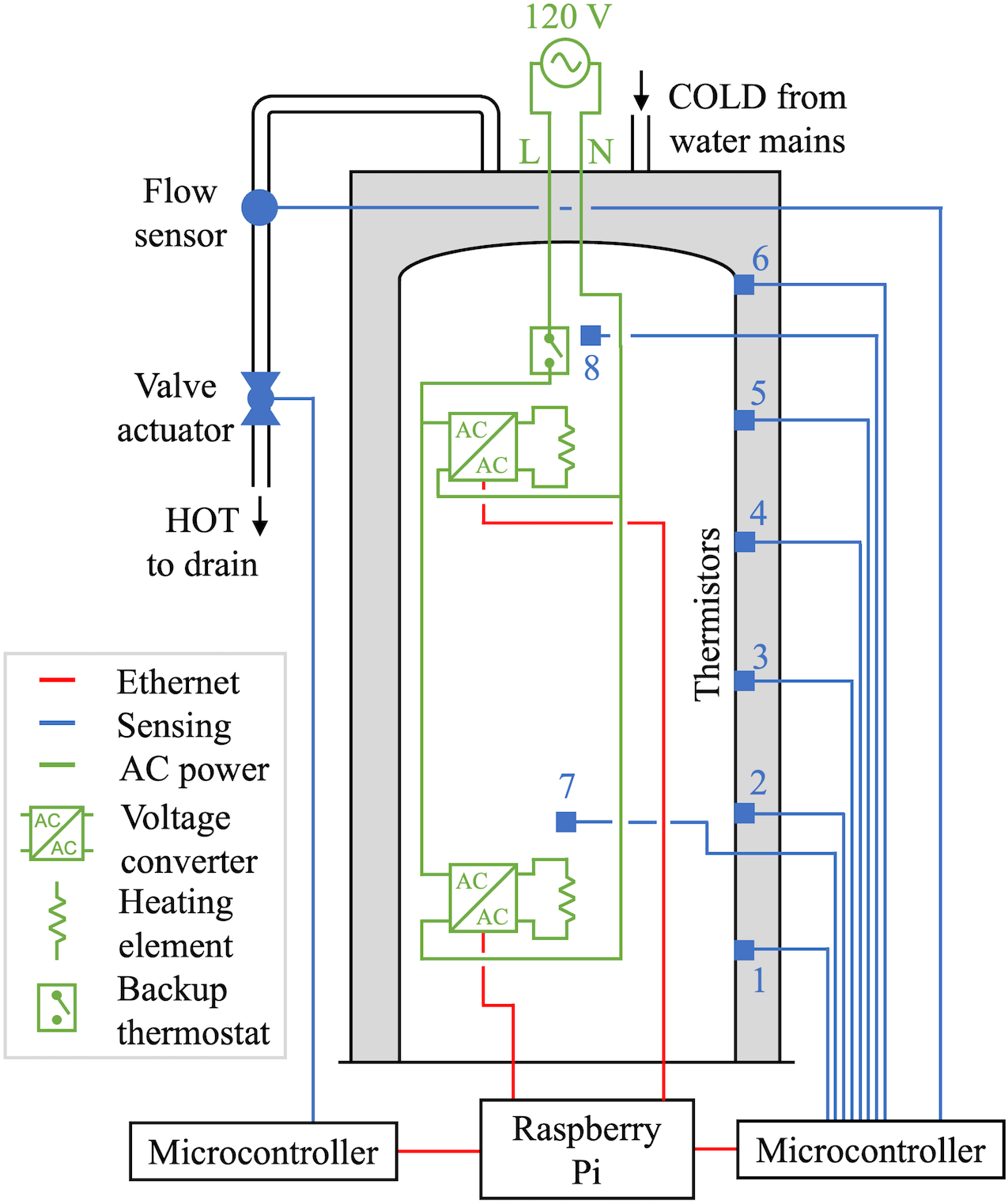}
\caption{Design of the laboratory water heater testbed.}\label{fig:testbed}
\end{figure}

Experiments were conducted on a 50 gallon Rheem Performance two-element resistive water heater (Fig. \ref{fig:testbed}). It was originally designed with two 4.5 kW elements operating non-simultaneously at 240 V with thermostatic hysteresis control. The elements were rewired for 120 V operation (1.13 kW per element) to be able to operate both elements simultaneously without exceeding the original water heater and circuit breaker current ratings. The water heater was also configured to be able to continuously control the power of each element. AC/AC voltage converters were used to modulate the voltage for each element to control the power \cite{goldin2017smart,goldin2022smart}. They are similar to commercially available ``solar diverter'' controllers for resistive water heaters which are used to maximize local PV self-consumption \cite{yildiz2021assessment}. Our lab setup allows the MPC control variables to represent actual consumed real power, rather than an average power that is converted to an ON/OFF signal. However, all control formulations can easily be implemented on resistive water heaters with only ON/OFF control capabilities. The results are expected to be nearly identical for short control intervals ($<$15 min) in terms of water temperature impacts and electricity costs. 

To ensure safety in case of communication or software errors, the original upper mechanical thermostat was used as a backup controller (Fig. \ref{fig:testbed}). It cuts power to both elements if the temperature rises above the upper limit of the deadband, which can be set significantly higher than $\bar{T}$. This backup controller never engaged during the tests conducted for this study.

The tank temperature was measured with thermistors that were installed on the outside of the inner metal tank by drilling holes through the outer metal casing and foam insulation (Fig. \ref{fig:testbed}). The measured values are similar to the internal water temperature and the approach resembles how water temperature is measured in many water heaters (e.g. with externally-mounted mechanical thermostats). Two sensors were placed near the original thermostats directly above each element (Fig. \ref{fig:testbed}, sensors 7 and 8), and six were placed on the side of the water heater at approximately equal spacing (Fig. \ref{fig:testbed}, sensors 1-6). Instead of directly measuring the outlet temperature, the sensor closest to the top of the tank (sensor 6) was used as a proxy for the outlet temperature during water draws. Sensor 7 was used for the one-node MPC, sensors 1, 7, and 8 were used for the three-node MPC, and sensors 7 and 8 were used for the thermostatic controller. 

The water heater was located in a temperature-controlled room with an ambient air temperature of approximately 70$^{\text{o}}$F. The inlet water generally measured between 64-70$^{\text{o}}$F during the laboratory testing period. Therefore in the MPC strategies, it was assumed that $T_a=70^\text{o}$F and $T_i=68^\text{o}$F.

The cumulative water flow was measured with a flow meter mounted on the outlet pipe with a measurement resolution of 0.05 gallons. Flow profiles were automated using a valve actuator installed in series with the flow meter. The flow rate was calibrated to the valve actuator position at a specific nominal water pressure ($\sim$ 55 psi), and an additional feedback controller was used to compensate for deviations in water pressure from the nominal value. This ensured that water draw profiles were highly reproduceable between tests despite large variations in building water pressure over time.

Microcontroller boards (TI EK-TM4C1294XL development boards) were used to record measurements of the element current, voltage, cumulative water flow volume, water temperatures, and valve actuator position at 1 Hz frequency. A Raspberry Pi (3 Model B+) was used for additional post-processing of measurements and for running the MPC optimization and thermostatic control logic. Data and commands were exchanged between the TI boards and Raspberry Pi with TCP/IP communication over an Ethernet connection. 

\subsection{Simulation-based Virtual 
Testbed}\label{sec:valid_sim}

While laboratory testing is important for validating control performance under realistic conditions, it is also very time consuming, as multi-day tests are generally required to capture daily patterns. A simulation-based virtual testbed based on a high-fidelity tank model that is tuned to real-world measurements can be useful for evaluating a larger number of scenarios and for tuning MPC hyperparameters.

For our simulation-based tests, we use a nonlinear multi-node model of tank thermal dynamics that is simulated in-the-loop with the MPC controller. The multi-node model is a one-dimensional PDE model \cite{xu2014modeling} that is spatially discretized. It accounts for temperature variation at different nodes along the vertical axis of the water heater \cite{xu2014modeling} and captures tank stratification more accurately than the three-node model. 

The dimensions of the lab water heater and positions of the elements were used to define the spatial characteristics of the multi-node model. Simulated state variables corresponding with the actual sensor positions in the lab water heater were used as sensor measurements in the simulation. 

Other model parameters were manually tuned using extensive data of water heater operation from laboratory tests under a variety of conditions. Results showed that at least 10 nodes were needed to optimize the temperature prediction accuracy (RMSE of temperature measurements). For all simulation-based results in this study, the number of nodes was set to 20, and the simulation timestep was set to 1 s, which allowed for accurate modeling of tank stratification while remaining computationally tractable and numerically stable.

\begin{figure}
\centering
\includegraphics[width=0.49\textwidth]{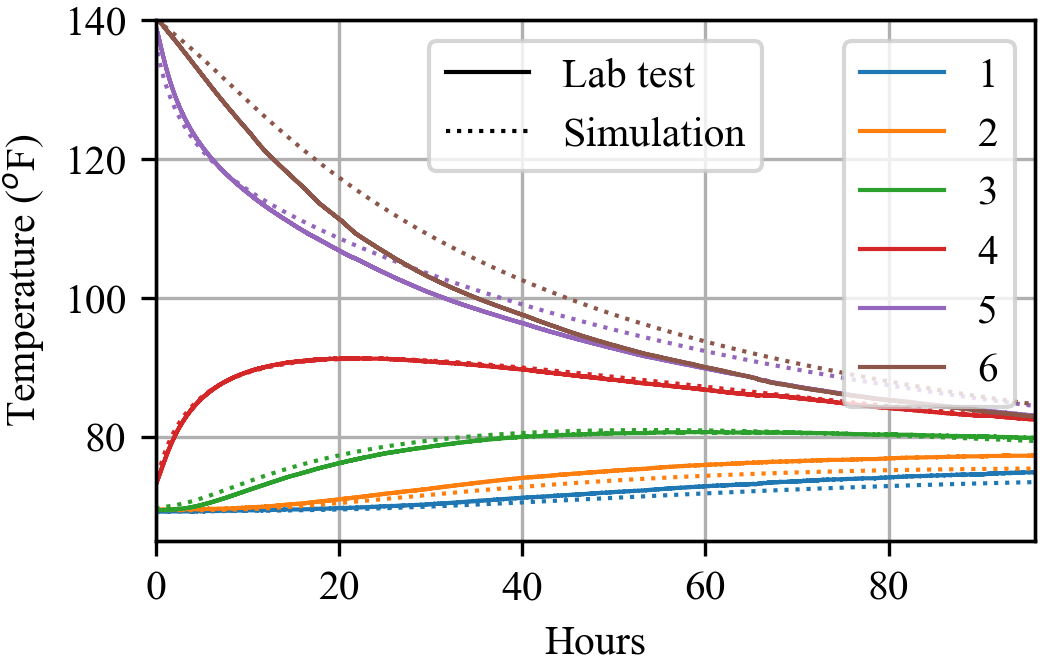}
\caption{Showing the similarity between lab testbed measurements and simulation results for the destratification of a water heater tank. Legend entries correspond with temperature sensor numbers, which are defined in Fig. \ref{fig:testbed}.}\label{fig:lab_sim_compare}
\end{figure}

Fig. \ref{fig:lab_sim_compare} shows an example comparison between measured and simulated temperatures when 20 nodes were used. The measurements are from sensors 1-6 in the laboratory testbed while the tank was destratifying. Before the beginning of the dataset, the top portion of the tank was heated to approximately 140$^\text{o}$F, while the bottom was left cold, creating significant tank stratification. Fig. \ref{fig:lab_sim_compare} shows how the tank destratified from this initial state over several days (with no water flow or element heating). The similarity between the measurements and simulation shows that the simulation accurately captures both ambient losses and conduction within the tank.

\section{Case Studies}\label{sec:case_studies}

In the following subsections we first show results for parameter identification of the controller models. Then we show results on controller performance under different scenarios and discuss the sensitivity to different factors such as sensor configuration and forecast errors. Finally we discuss the computation time of each controller implementation.

\subsection{Parameter Identification}\label{sec:param_id_results}

\subsubsection{One-node model}

We evaluate the parameter identification method described in Section \ref{sec:param_id} using measurements from the lab testbed. The datasets described in Section \ref{sec:param_id} were collected under two different starting conditions: (i) when the tank was initially well-mixed, and (ii) when the tank was initially stratified. Table \ref{tab:1node_param} shows the parameter identification results for $\bar{\delta t}=$ 300 s. For the well-mixed case, the estimated volume is fairly close to the actual tank volume that is above the lower element, which is the volume that is directly heated when the element is on. The estimate for $U$ is also similar in magnitude to values from other studies for the one-node model. However, for the stratified case, the estimates for $V$ and $U$ are very different than the well-mixed case and are not physically realistic. Therefore, for the one-node model it is important that all model assumptions hold in the dataset that the model is trained on to obtain physically realistic parameter estimates.

\begin{table}[]
\caption{Identified parameters for one-node model}
    \centering
    \begin{tabular}{|c|c|c|}
      \hline
      Parameter   & Well-mixed tank & Stratified tank \\
      \hline
       $V$  & 0.156 & 0.069 \\
       \hline
       $U$  & 1.27 & 11.1 \\
        \hline
    \end{tabular}
    \label{tab:1node_param}
\end{table}

\begin{table}[]
\caption{Identified parameters for three-node model}
    \centering
    \begin{tabular}{|c|c|}
      \hline
      Parameter   & Value \\
      \hline
       $U_u$  & 0.662 \\
       \hline
       $U_m$  & 0.092 \\
        \hline
       $U_l$  & 1.15 \\
        \hline
       $V_u$  & 0.0546 \\
        \hline
    \end{tabular}
    \quad \quad
    \begin{tabular}{|c|c|}
      \hline
      Parameter   & Value \\
      \hline
       $V_m$  & 0.0932 \\
        \hline
       $V_l$  & 0.0415 \\
        \hline
       $K_{ml}$  & 3.59 \\
        \hline
       $K_{um}$  & 0.703 \\
        \hline
    \end{tabular}
    \label{tab:3node_param}
\end{table}

\subsubsection{Three-node model}

The three-node parameter identification procedure described in Section \ref{sec:param_id} was also tested using data from the lab testbed. Table \ref{tab:3node_param} shows the parameter identification results. The three identified volumes are relatively close to the actual tank volumes that are above the upper element, between the two elements, and below the lower element. The results for the other parameters are within the range of physically realistic values.

\subsection{Control Performance Metrics}

We quantify controller performance in terms of: (i) the total electricity cost [$\$$], (ii) the average cost per kWh of electricity, and (iii) the average cost per kWh of water consumed. The last metric normalizes for the amount of energy embodied in the water drawn from the tank and consumed by the user, which can vary slightly under different controllers. The energy embodied in hot water draws [kWh] for a certain period is defined as:
\begin{align}
    E_{d}=\frac{1}{3.6e6}\sum_{j=0}^{N_{s}}{\left[c_p \rho V_{use}(\tau_j)(T_{out}(\tau_j)-T_{i})\right]}
\end{align}

where $V_{use}(\tau_j)$ is the volume of water drawn between $\tau_j$ and $\tau_{j+1}$. $T_{out}$ is the outlet temperature $[K]$, which is estimated using the top temperature sensor for lab tests (sensor 6, Fig. \ref{fig:testbed}) and the top node temperature for the multi-node simulations. $N_s$ is the number of timesteps in the evaluation period. The factor in the denominator accounts for unit conversion.

We also analyze the distribution of water draw temperatures for each controller to evaluate comfort impacts.

\begin{figure*}\centering
\includegraphics[width=0.99\textwidth]{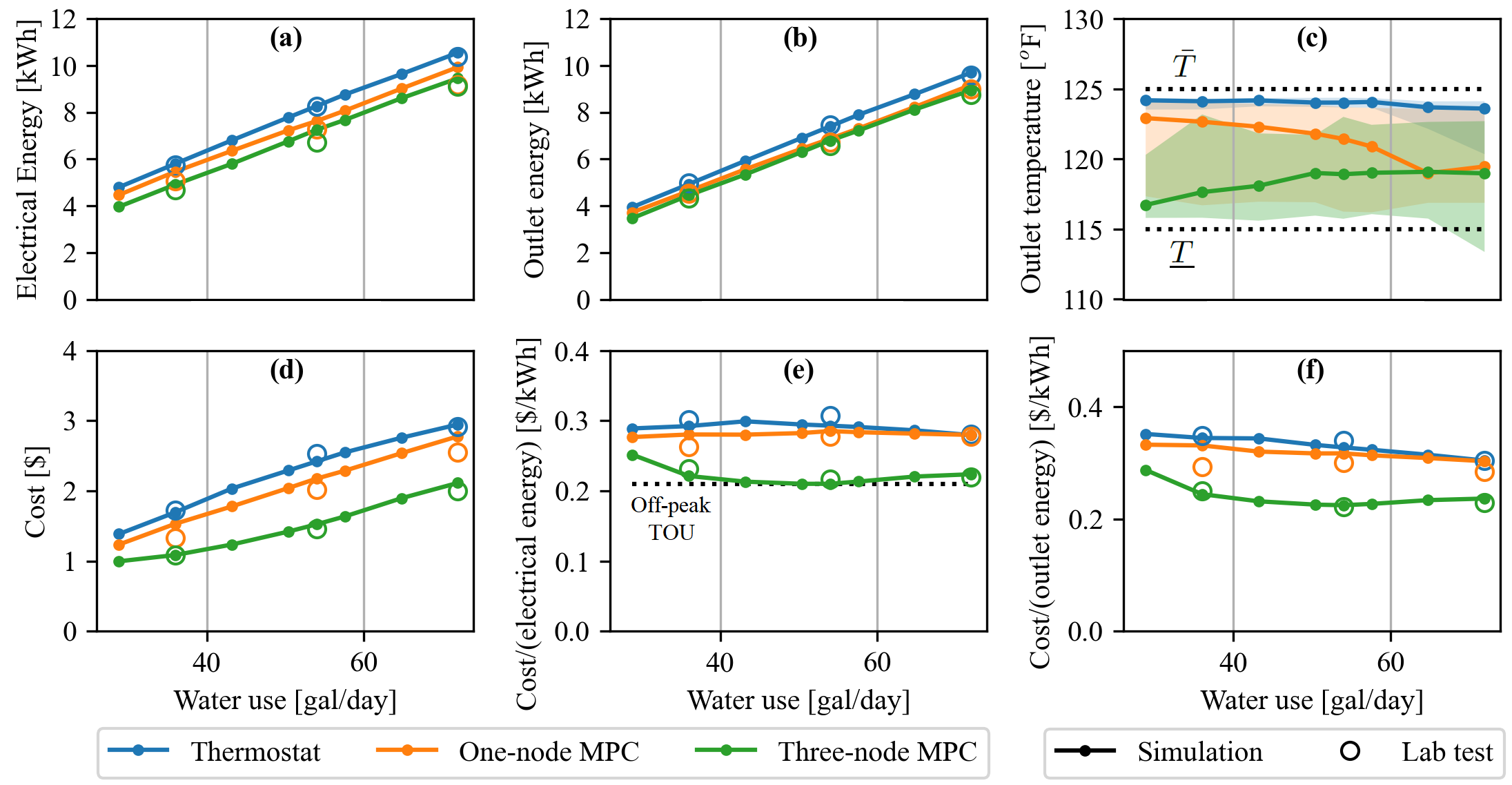}
\caption{Results from lab tests and simulations showing: (a) electrical energy consumption [kWh], (b) energy embodied in drawn water [kWh], (c) outlet water temperature [$^\text{o}$F], (d) cost [\$], (e) cost per kWh of electrical energy [\$/kWh], and (f) cost per kWh of energy embodied in drawn water [\$/kWh], as a function of daily water usage [gal/day]. In plot (c), the shaded area represents range between the 10th and 90th percentile in the simulation-based tests and the solid line represents the mean temperature. Energy consumption is similar between controllers, while costs are significantly reduced for the three-node MPC relative to the one-node MPC and the thermostatic controller.}\label{fig:results}
\end{figure*}

\subsection{Perfect Foresight of Water Use Patterns}

First, we compare the performance of the three different controllers under a simplified scenario: with perfect foresight of future hourly average water use patterns. Hourly average flow rates are calculated from the actual flow profile that is actuated on the water heater. The two MPC controllers (one-node and three-node) use those average hourly flow rates as a forecast. Therefore, the forecast received by the MPC controller is accurate in its general daily shape and magnitude, but is not as granular as the actual draw profile. The thermostatic controller does not use water use forecasts. Considering this perfect foresight case is useful because it allows us to to isolate the effects of plant-model mismatch on MPC performance, while removing other sources of uncertainty. Other scenarios with forecast errors are explored in Section \ref{sec:uncertainty}.

These tests were conducted on both the lab water heater and the simulation-based testbed. For water flow patterns, we constructed a synthetic daily flow profile that has similar characteristics to those observed in field studies \cite{lutz2014typical,ecotope2015} in terms of timing, flow volume, and flow rate. The profile has a granularity of 30 s. Water usage is primarily clustered during the morning and evening times, including during the TOU peak price period. A base profile was constructed with a daily consumption of 36 gal/day, and then additional profiles were created by scaling the duration of each draw, keeping the timing and flow rate the same. The same 24 hr water draw pattern was repeated each day of the test. Lab tests were conducted for each controller with 36, 54, and 72 gal/day flow profiles. Simulation-based tests were conducted for eight different daily water usages ranging from 28.8 to 72 gal/day.

Tests were performed using a TOU cost profile with a peak price of $\$0.47$/kWh from 5-8pm and a off-peak price of $\$0.21$/kWh at all other times. Each lab and simulation-based test was run for three days. At the beginning of each test, the tank was initialized by heating the water above the lower element to 120$^\text{o}$F. Performance metrics were calculated on only the final day of the test to allow the controller to reach a steady-state behavior, since the temperature initialization at the beginning of the test can affect the results. MPC hyperparameters were selected based on preliminary simulation results, and are listed in Table \ref{tab:mpc_params}. The thermostatic controller was updated every 30~s. Parameters in Table \ref{tab:1node_param} (well-mixed case) and Table \ref{tab:3node_param} were used for the one-node and three-node MPC, respectively.

\begin{table}[]
\caption{MPC hyperparameters used for case studies}
    \centering
    \setlength\extrarowheight{2.5pt}
    \begin{tabular}{|c|c|}
      \hline
      Parameter   & Value \\
      \hline
       $\delta t$  & 600 s \\
       \hline
       $\bar{\delta}t$  & 300 s \\
        \hline
    \end{tabular}
    \quad
        \begin{tabular}{|c|c|}
      \hline
      Parameter   & Value \\
        \hline
        $T_N$ & 18 h\\
        \hline
        $N$ & 108\\
        \hline
    \end{tabular}
    \quad
            \begin{tabular}{|c|c|}
      \hline
      Parameter   & Value \\
        \hline
        $\underline{T}$ & $115^\text{o}$F\\
        \hline
        $\bar{T}$ & $125^\text{o}$F\\
        \hline
    \end{tabular}
    \label{tab:mpc_params}
\end{table}

\subsubsection{Energy and comfort impacts}
Fig. \ref{fig:results} shows different performance metrics and summary statistics as a function of daily water usage. Overall, lab tests and simulation-based tests yield similar results.  Differences can be attributed to small deviations between actual and simulated thermal dynamics. As shown in Fig. \ref{fig:results}a and \ref{fig:results}b, the total energy consumption and total energy embodied in water draws were very similar across all three controllers. The consumption of the three-node MPC controller is slightly lower than the other two controllers since it can keep the temperature of drawn hot water closer to the bottom of the comfort deadband. This very slightly decreases the amount of energy that is embodied in the water drawn from the tank. It also slightly reduces ambient losses, improving overall energy efficiency by a small amount. This can also be observed in Fig. \ref{fig:results}c, which shows the mean water draw temperature for each controller from simulation-based results, as well as the range between the 10th and 90th percentile temperatures. For all controllers, most water draws occur within the comfort band between $\underline{T}$ and $\bar{T}$.

\subsubsection{Electricity costs and load shifting}
In terms of electricity costs, the three-node MPC formulation performs significantly better than both the one-node MPC and thermostatic controller (Fig. \ref{fig:results}d). For laboratory tests, results show cost reductions of 12.3-23.2\% for the one-node MPC and 31.2-42.5\% for the three-node MPC relative to the thermostatic controller. Most of these cost reductions are driven by a shift in load from on-peak to off-peak pricing periods, as opposed to improvements in energy efficiency. The three-node MPC is better able to shift load from the peak period to the off-peak period than the one-node MPC. This can also be observed in Fig. \ref{fig:results}e, which shows the average cost of consumed electricity under each controller. For the three-node MPC, the average cost per kWh is close to the off-peak price (\$0.21/kWh), meaning that almost all load is shifted to that period. Fig. \ref{fig:results}f shows that even after normalizing for slight differences in total energy consumption between controllers, the three-node MPC results in significant cost savings. Since the three-node MPC is already able to shift almost all load from on-peak to off-peak periods, additional improvements to the controller (e.g., use of a higher-fidelity control model) are expected to result in fairly small additional improvements in performance.

\begin{figure*}\centering
\includegraphics[width=0.99\textwidth]{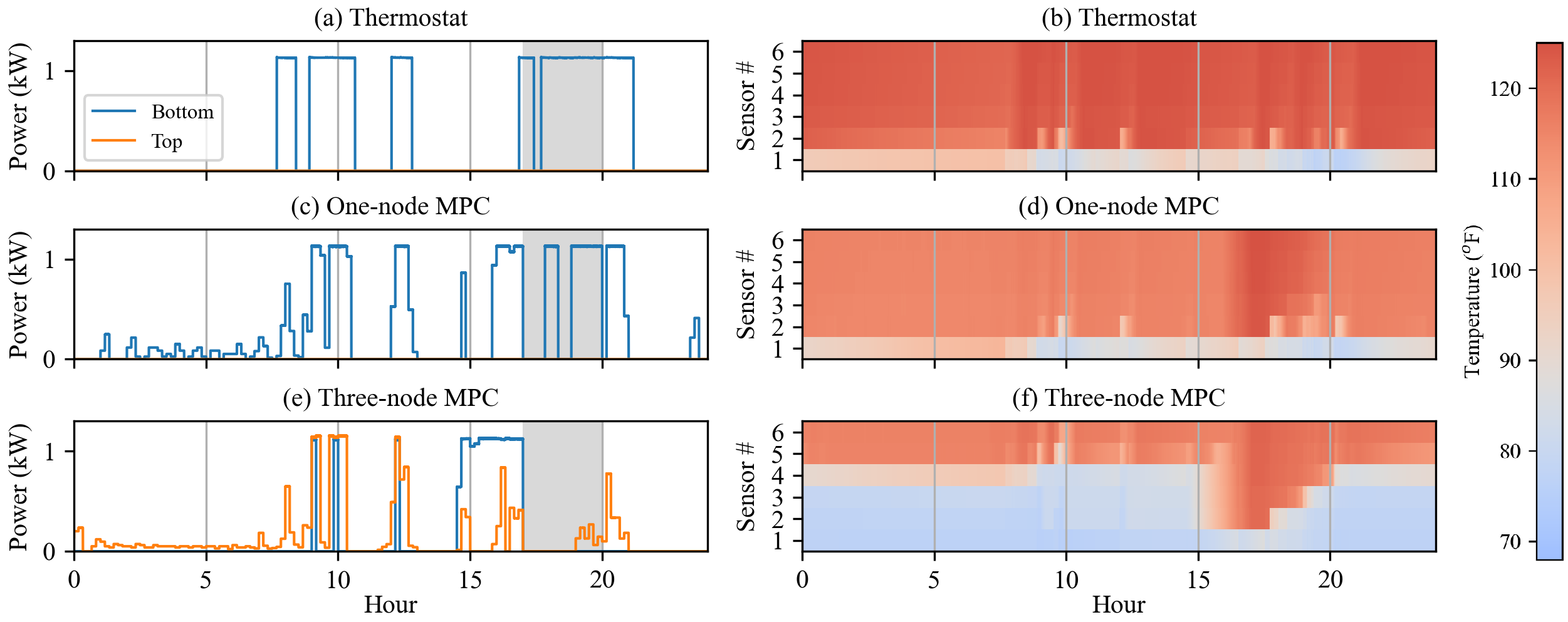}
\caption{Power consumption profiles and tank temperature heat map for the thermostat controller, one-node MPC, and three-node MPC over a day of laboratory tests for the 54 gal/day water usage case. The TOU peak-price window is shaded in gray. In subplots (b), (d), and (f), the sensor numbers correspond with the sensor numbers in Fig. \ref{fig:testbed}. A shown, the three-node MPC is better able to shift load out of the peak price window.} \label{fig:timeseries}
\end{figure*}

To understand why the three-node MPC results in more effective cost reduction than the one-node MPC, it is useful to analyze controller behavior over time. Fig. \ref{fig:timeseries} shows the power consumption and tank temperature profiles of the three controllers for a single day of laboratory tests. The thermostatic controller tends to consume power after large water draws or after long periods with no power consumption. In the case shown, the lower element is on for most of the peak price period. The one-node MPC does preheat the tank before the peak price period, however it reactively turns on during the peak period after water draws, even though there is enough hot water in the tank. This reactive behavior is driven by the large mismatch between the controller model and the actual tank dynamics. The one-node model, which assumes the tank is well-mixed, predicts a small, smooth decrease in temperature during a water draw. However, real-world stratification produces a larger, sharp decrease in measured water temperature, particularly if the sensor is placed near the bottom of the tank. The measured temperature can fall significantly below $\underline{T}$, causing the one-node MPC to reactively turn on the element, regardless of the amount of hot water at the top of the tank. This behavior is highly undesirable for load shifting applications, as the goal is to shift as much load as possible away from the peak price period, when a significant portion of water usage may occur. Because the three-node model more accurately models tank stratification, the controller does not reactively turn on the element during large draws and instead allows the bottom of the tank to become cold, when possible. Additionally, with two heating elements to control, the three-node MPC has an extra degree of freedom over controlling the tank temperature profile compared to the one-node MPC, which we assume can only utilize one element. The lower element must heat up a much larger volume of water to affect the outlet water temperature compared to the upper element. These differences can affect load shifting capabilities.

We also analyzed whether the one-node MPC performance could be improved by instead using multiple temperature sensors to measure the average tank temperature. In simulation, we tested using the average of five temperature sensors (sensors 2-6 in Fig. \ref{fig:testbed}) as the state variable. Performance did improve, however it was still significantly worse than the performance of the three-node MPC, for most scenarios.

For the one-node MPC, we assume that only the lower element is used. Therefore, the maximum power consumption is half of that of the three-node MPC. We analyzed whether this assumption could be contributing to the poor performance of the one-node MPC by rerunning the simulation results  with the maximum power of the lower element $\bar{p}$ set to be twice as large. However, the results were very similar to the original case, and sometimes even worse. This suggests that this assumption is not a primary factor driving the poor performance. 

\subsection{Performance with Water Forecast Errors}\label{sec:uncertainty}

\begin{figure}
\centering
\includegraphics[width=0.48\textwidth]{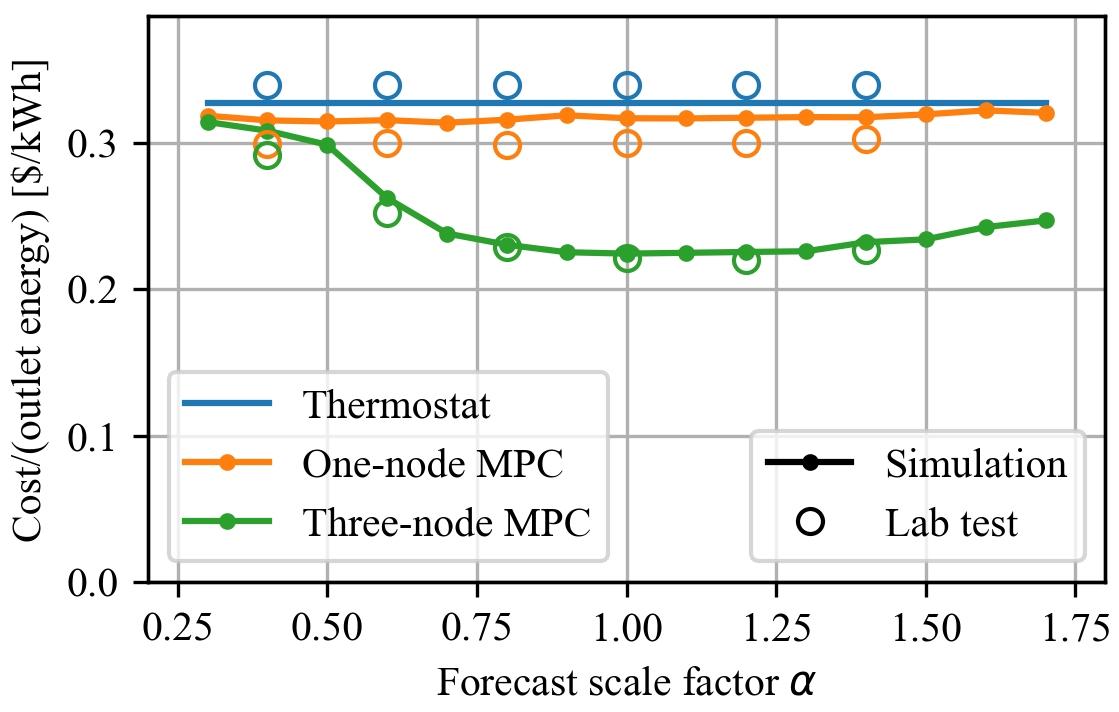}
\caption{Controller performance (cost per kWh of drawn hot water) as a function of the forecast scale factor $\alpha$, for the 54 gal/day actual draw profile. The three-node MPC outperforms the one-node MPC and thermostat controller over a wide range of forecast errors.}\label{fig:uncertainty}
\end{figure}

Since there is significant uncertainty and variability in real-world water use patterns, the sensitivity of each controller to forecast errors should be investigated. However, there is currently little open-source, publicly available, long-term data on residential water flow patterns, which makes it challenging to perform such analysis without making large assumptions. Instead, we provide a simple analysis using both lab tests and simulation-based tests that show how controller performance changes when daily water consumption is significantly underpredicted and overpredicted. A more detailed uncertainty analysis is left to future work.  

We simulate forecast errors by scaling the actual hourly average flow rates by a constant factor $\alpha$. For example, for $\alpha=1.5$, the MPC controller receives predictions that are 50\% too large for all hours in the prediction horizon. Therefore, the daily shape of the predicted flow profile is accurate, but the magnitude is significantly inaccurate. For small and large values of $\alpha$, this represents fairly extreme examples of forecast errors. We vary $\alpha$ between 0.3 and 1.7 for a given actual flow profile and analyze how performance changes. The thermostatic controller does not use water forecasts, so the performance does not depend on $\alpha$. Therefore, only a single simulation and lab test was conducted for the thermostatic controller.

Results are included in Fig. \ref{fig:uncertainty} for the 54 gal/day actual flow profile, showing how the cost per kWh of drawn water varies as a function of $\alpha$. The three-node MPC performs significantly better than both the one-node MPC and thermostatic controller in terms of cost savings over a very large range of $\alpha$ values. Overpredicting water use has a very small impact on cost, while large under-predictions can significantly degrade performance. Large underpredictions can also result in thermal comfort impacts.

\subsection{Computation Time}

A commonly cited disadvantage of MPC relative to RL or rule-based control strategies is the increased computational complexity \cite{ruelens2016reinforcement}. In this work, the average computation time to solve one optimization problem on a Raspberry Pi was 0.283 s for the one-node MPC and 0.635 s for the three-node MPC. These solution times are more than two orders of magnitude faster than the MPC control interval $\delta t$. This demonstrates that such strategies can easily be implemented in real-time using open-source solvers on a low cost microcomputer.

\section{Conclusions and Future Work}\label{sec:con}

In this work we analyzed how modeling tank stratification in water heater MPC strategies affects controller performance. We proposed a three-node MPC formulation and compared it to a thermostatic controller and a one-node formulation for load shifting under TOU rates, using both real-time laboratory testing and simulation-based analysis. Results indicate that the one-node model results in very minimal cost savings over thermostatic control, due to significant plant-model mismatch. However the proposed three-node MPC formulation, which explicitly accounts for tank stratification, achieves appreciably greater load shifting and cost savings. 

The approaches developed in this paper could be extended to other water heater control objectives (e.g., carbon emissions, real-time prices) or water heater designs. Heat pump water heaters (HPWHs) are becoming more common due to their energy efficiency. However, they are different from electric resistance water heaters in various ways which would need to be accounted for in an MPC strategy (e.g., compressor cycling constraints, relationship between coefficient of performance and water temperature). Unlike resistive water heaters, the design of HPWHs varies significantly between different manufacturers \cite{sparn2014laboratory}. Therefore, control strategies may need to be tailored to certain models/manufacturers. The use of thermostatic mixing valves on water heaters is also becoming increasingly common, as they can enable greater load shifting capabilities. However, mixing valves introduce more complexity into MPC formulations, which must be considered. Additionally, future work could also analyze the effects of water use uncertainty in greater detail, using real-world data. A stochastic MPC formulation based off of the proposed stratified tank model could potentially further increase performance and controller robustness. Finally, in this work we focused on a local control objective (i.e., minimizing the electricity costs for a single consumer). Future research could analyze the impacts of the proposed strategy on aggregate-level load profiles and consider how the proposed work could be extended to aggregate-level control applications (e.g., peak load management).

\bibliographystyle{IEEEtran}
{\bibliography{ref}}
\end{document}